\documentclass[review,3p,authoryear]{elsarticle}

\usepackage{amssymb}                                            
\usepackage{amsmath}
\usepackage{graphicx}
\usepackage{url}	
\usepackage[hidelinks]{hyperref} 

\usepackage{lineno}

\hypersetup{
 colorlinks=true,
 allcolors=black,
 citecolor=blue
 }

\journal{Elsevier}

\begin{document}
\begin{frontmatter}

\title{Including the effect of depth-uniform ambient currents on waves in a non-hydrostatic wave-flow model}
\author[add1]{Dirk P. Rijnsdorp\corref{cor1}}

\ead{d.p.rijnsdorp@tudelft.nl}
\cortext[cor1]{Corresponding author.}

\author[add2]{Arnold van Rooijen}
\author[add1]{Ad Reniers}
\author[add1]{Marion Tissier}
\author[add1,add3]{Floris de Wit}
\author[add1]{Marcel Zijlema}

\address[add1]{Environmental Fluid Mechanics section, Faculty of Civil Engineering and Geosciences, Delft University of Technology, Netherlands}
\address[add2]{Oceans Graduate School \& UWA Oceans Institute, The University of Western Australia, Australia}
\address[add3]{Svasek Hydraulics, The Netherlands}

\begin{linenomath*}
\begin{abstract}
Currents can affect the evolution of waves in nearshore regions through altering their wavenumber and amplitude. Including the effect of ambient currents (e.g., tidal and wind-driven) on waves in phase-resolving wave models is not straightforward as it requires appropriate boundary conditions in combination with a large domain size and long simulation duration. In this paper, we extended the non-hydrostatic wave-flow model SWASH with additional terms that account for the influence of a depth-uniform ambient current on the wave dynamics, in which the current field can be taken from an external source (e.g., from observations or a circulation model). We verified the model ability by comparing predictions to results from linear theory, laboratory experiments and a spectral wave model that accounts for wave interference effects. With this extension, the model was able to account for current-induced changes to the wave field (i.e., changes to the wave amplitude, length and direction) due to following and opposing currents, and two classical examples of sheared currents (a jet-like current and vortex ring). Furthermore, the model captured the wave dynamics in the presence of strong opposing currents. This includes reflections of relatively small amplitude waves at the theoretical blocking point, and transmission of breaking waves beyond the theoretical blocking point for larger wave amplitudes.  The proposed model extension allows phase-resolving models to more accurately and efficiently simulate the wave dynamics in coastal regions with tidal and/or wind-driven flows.
\end{abstract}
\end{linenomath*}

\begin{keyword}
wave-current interactions, non-hydrostatic, SWASH.
\end{keyword}

\end{frontmatter}

\section{Introduction}\label{sec:Introduction}
Complex coastal regions such as estuaries and tidal inlets often feature the joint occurrence of surface gravity waves (e.g., swell and wind seas) and currents (e.g., riverine, tidal, and wind-driven flows). These processes typically occur at different spatial and temporal length scales. Currents generally experience variations at hour to day timescales and over $\mathcal{O}(\mathrm{km})$ length scales. To the contrary, waves have periods of several seconds and length scales of $\mathrm{O}(10-100\;\mathrm{m})$. 

Waves propagating over spatially varying currents conserve wave action \citep[e.g.,][]{Bretherton1968WavetrainsMedia,Mei2005} but experience a change in their wavelength associated with the Doppler' shift \citep[e.g.,][]{Peregrine1976InteractionCurrents,Holthuijsen2007}. As a result, the wave celerity and group velocity change, resulting in changes in wave amplitude and wave direction (current-induced shoaling and refraction). In strong currents that oppose the direction of wave propagation, the group velocity $c_g$ approaches zero, resulting in significant increases of the wave height and wave-blocking when $c_g=0$ \citep[e.g.,][]{Chawla2002MonochromaticPoints}. Furthermore, waves steepen in opposing currents which may trigger wave breaking resulting in additional dissipation of wave energy \citep[e.g.,][]{Chawla2002MonochromaticPoints}. Current-induced changes in the wave shape can in turn impact the magnitude of wave-driven sediment transport \citep[e.g.,][]{Roelvink1989Bar-generatingBeach,Hoefel2003Wave-InducedMigration}. Including for the current effects on waves is thus important when predicting sediment transport and the resulting morphological changes in coastal regions.

To date, modelling of combined wave-current actions in coastal regions has generally relied on the coupling of phase-averaged wave models and circulation models \citep[e.g.,][]{Lesser2004DevelopmentModel,Roelvink2009,Uchiyama2010,Kumar2012,Dodet2013Wave-currentInlet,Olabarrieta2014TheAnalysis} through either the radiation stress \citep[e.g.,][]{LHS1962,LHS1964} or vortex force formalism \citep[e.g.,][]{Craik1976,McWilliams2004}. Such coupled models have been successfully adopted to simulate the hydrodynamics in a variety of nearshore regions, ranging from sandy beaches \citep[e.g.,][]{Orzech2011MegacuspsModeling,Hansen2015,Luijendijk2017TheStudy,Rafati2021ModelingEvents} to tidal inlets and rivers where strong ambient currents can occur \citep[e.g.,][]{Dodet2013Wave-currentInlet,Chen2015HydrodynamicWaves,Nienhuis2016AlongshoreMorphodynamics,Hopkins2018StormInlet}. However, such a coupled approach relies on spectral wave models that do not intrinsically account for phase-dependent (e.g., wave-interference and diffraction) and nonlinear wave processes (e.g., triad interactions and wave breaking) but rely on parametrizations thereof.


As an alternative to phase-averaged wave models, phase-resolving wave models have been developed to simulate the nearshore evolution of waves in the presence of ambient currents. Linear phase-resolving wave models based on the mild-slope equations have been shown to capture changes to the wave kinematics associated with the Doppler shift \citep[e.g.,][]{Booij1981,Kirby1986}. This has allowed such models to capture the effect of prescribed ambient currents on the nearshore wave evolution \citep[e.g.,][]{Chen2005OnEquation,Touboul2016ExtendedCurrent}. Models based on the mild-slope equations generally rely on assumptions of linear wave theory, although they can be extended to account for higher order wave effects \citep[e.g.,][]{Kaihatu1995}. Furthermore, they do not inherently account for wave-induced currents but require a coupling to a circulation model to capture such effects.

Alternatively, weakly to fully nonlinear phase-resolving wave-flow models based on Boussinesq-type formulations \citep[e.g.,][]{Peregrine1967,Madsen1991,Nwogu1993,Kirby2016BoussinesqScales} or the non-hydrostatic approach \citep[e.g.,][]{Zijlema2011,Ma2012,Wei2014} can be used to simulate waves and wave-induced currents in coastal regions \citep[e.g.,][]{Chen1999,Feddersen2011,Rijnsdorp2015a,Baker2021ModeledBeach}. Such models intrinsically account for phase-dependent wave effects, nonlinear wave interactions, and the generation of wave-induced currents (e.g., longshore currents and rip currents).
However, directly including tidal and/or wind-driven currents in such models is not straightforward due to the range of spatial and temporal scales required. For example, including tidal currents in a phase-resolving model would typically require a significantly larger computational time to allow for spin-up of the tidal flow and a larger domain with appropriate boundary conditions to allow for the propagation of the tidal wave in and out of the domain. Due to the excessive computational costs of such a model setup, this presently inhibits a direct inclusion of such currents in phase-resolving wave-flow models.

Several efforts have been made to account for the interactions between waves and a prescribed ambient current in nonlinear phase-resolving models based on the Boussinesq or non-hydrostatic approach. Most efforts focused on extending Boussinesq-type formulations to account for interactions between waves and an ambient current \citep[e.g.,][]{Son2014InteractionProfile,Yang2020Depth-integratedApplications,Yang2022Depth-integratedProfile}. Efforts to extend non-hydrostatic models have been limited to \citet{deWit2017IncludingModel}, who added a spatially homogeneous pressure term in the alongshore momentum equation of a non-hydrostatic model to simulate the nearshore wave dynamics in the presence of alongshore tidal flows at a sandy beach.
Despite this progress on including the effect of ambient currents on waves in nonlinear phase-resolving wave-flow models, their application at complex coastal sites have not yet been able to account for the effect of spatially varying current fields from tides and/or wind on the wave dynamics \citep[e.g.,][]{Risandi2020HydrodynamicModel,Rijnsdorp2021ASystem,Baker2021ModeledBeach}.

In this work, we extend the non-hydrostatic wave model SWASH \citep{Zijlema2011} to account for the effect of a prescribed depth-uniform ambient current on the wave dynamics, in which the current field can be obtained from an external source (e.g., observations or a circulation model). By introducing a separation of scales and assuming vertically uniform mean flows, we derive additional terms to the governing equations that account for the effect of a spatially varying depth-uniform current on the waves (Section \ref{sec:NumMethod}). Comparisons with linear wave theory, a spectral wave model and flume experiments show that the proposed model is able to account for changes in the wave height and wavelength due to an ambient currents (Section \ref{sec:LinProp}-\ref{sec:TC}). In Section \ref{sec:Discussion}-\ref{sec:Conclusions}, we conclude that the proposed extension allows non-hydrostatic models to account for the effect of ambient currents on waves.


\section{Numerical Methodology}\label{sec:NumMethod}

\subsection{Governing equations}
The governing equations of the model are the Reynolds-Averaged Navier-Stokes (RANS) equations for an incompressible fluid that is bounded by the bottom $d(x,y)$ and a free-surface $\zeta(x,y,t)$, where ($x,y,z$) are the Cartesian coordinates and $t$ is time,
\begin{linenomath}
\begin{align}
&\frac{\partial u}{\partial x} + \frac{\partial v}{\partial y} + \frac{\partial w}{\partial z} = 0, \\
&\frac{\partial u}{\partial t} + u \frac{\partial u}{\partial x} + v \frac{\partial u}{\partial y} + w \frac{\partial u}{\partial z} + g \frac{\partial \zeta}{\partial x} + \frac{\partial p_{nh}}{\partial x} = \frac{\partial \tau_{xx} }{\partial x} + \frac{\partial \tau_{xy}}{\partial y} + \frac{\partial \tau_{xz}}{\partial z},\\
&\frac{\partial v}{\partial t} + u \frac{\partial v}{\partial x} + v \frac{\partial v}{\partial y} + w \frac{\partial v}{\partial z} + g \frac{\partial \zeta}{\partial y} + \frac{\partial p_{nh}}{\partial y} = \frac{\partial \tau_{yx} }{\partial x} + \frac{\partial \tau_{yy}}{\partial y} + \frac{\partial \tau_{yz}}{\partial z},\\
&\frac{\partial w}{\partial t} + u \frac{\partial w}{\partial x} + v \frac{\partial w}{\partial y} + w \frac{\partial w}{\partial z} + \frac{\partial p_{nh}}{\partial z} = \frac{\partial \tau_{zx} }{\partial x} + \frac{\partial \tau_{zy}}{\partial y} + \frac{\partial \tau_{zz}}{\partial z}.
\end{align}
\end{linenomath}
In this set of equations, $p_{nh}$ is the non-hydrostatic pressure, ($u,v,w$) are the velocity components in ($x,y,z$) direction, respectively, $\tau$ represents the turbulent stress (estimated using an eddy viscosity approximation). The kinematic boundary conditions at the bottom and the free-surface follow from the assumption that the vertical boundaries of the fluid are single valued functions of the horizontal coordinates,
\begin{linenomath}
\begin{align}
    &w_{z=\zeta}=\frac{\partial \zeta}{\partial t} + u \frac{\partial \zeta}{\partial x} + v \frac{\partial \zeta}{\partial y}, \\
    &w_{z=-d}= - u \frac{\partial d}{\partial x} - v \frac{\partial d}{\partial y}.
\end{align}
\end{linenomath}
Integrating the local continuity equation over the water column results in a global continuity equation that describes the temporal evolution of the free-surface,
\begin{linenomath}
\begin{equation}
    \frac{\partial \zeta}{\partial t} + \frac{\partial}{\partial x} \int\limits_{-d}^{\zeta} u \mathrm{d}z + \frac{\partial }{\partial y} \int\limits_{-d}^{\zeta} v \mathrm{d}z=0.
\end{equation}
\end{linenomath}
Assuming a constant atmospheric pressure (equal to zero for convenience) and neglecting viscous stresses at the free-surface, the non-hydrostatic pressure is set to zero at the free-surface \citep[e.g.,][]{Stelling2003}. At the bottom, the tangential stress is prescribed based on the quadratic friction law (in the case of a coarse vertical resolution) or the law of the wall (in the case of a fine vertical resolution). Turbulent stresses are modelled using the eddy-viscosity model and the k-$\epsilon$ turbulence closure model \citep[See][for more details]{Rijnsdorp2017}. Combined with boundary conditions at all horizontal edges of the physical domain, the above set of equations forms the basis of the SWASH model. 

\subsection{Including the effect of currents on waves}
In this work, we set out to decouple the modelling of the surface waves and the currents that are slowly-varying with respect to the wave timescale (e.g., tidal currents and wind-driven currents). With this approach, we aim to account for the current effect on waves through prescribing an ambient current field from an other model (e.g., a circulation model) that alters the wave dynamics solved by the RANS equations.

To this end, we separate the horizontal flow variables and surface elevation as,
\begin{linenomath}
\begin{align}
    u(x,y,z,t) = U(x,y) + u^\prime(x,y,z,t),\\
    v(x,y,z,t) = V(x,y) + v^\prime(x,y,z,t),\\
    \zeta(x,y,t) = \eta(x,y) + \zeta^\prime(x,y,t).
\end{align}
\end{linenomath}
In these equations, ${[...]}^\prime$ denotes variables which we associate with wave-related motions and wave-induced currents. Capital letters ($U$ and $V$) represent vertically uniform horizontal flow velocities and $\eta$ a mean water level, which both vary over a timescale much larger than the wave motions and are considered to be constant over the wave-timescale. Substituting this separation of variables into the governing equations and neglecting the viscous contributions and tangential stress at the bottom yields,
\begin{linenomath}
\begin{align}
&\frac{\partial U+u^\prime}{\partial x}+\frac{\partial V+v^\prime}{\partial y} + \frac{\partial w}{\partial z} = 0,\label{eq:LCi}\\
&\frac{\partial U+u^\prime}{\partial t}+(U+u^\prime)\frac{\partial U + u^\prime}{\partial x} + (V+v^\prime)\frac{\partial U+u^\prime}{\partial y} + w \frac{\partial U+u^\prime}{\partial z} + g \frac{\partial \eta+\zeta^\prime}{\partial x} + \frac{\partial p_{nh}}{\partial x} = 0,\label{eq:momUi}\\
&\frac{\partial V+v^\prime}{\partial t} + (U+u^\prime)\frac{\partial V+v^\prime}{\partial x} + (V+v^\prime)\frac{\partial V + v^\prime}{\partial y} + w \frac{\partial V+v^\prime}{\partial z} + g \frac{\partial \eta+\zeta^\prime}{\partial y} + \frac{\partial p_{nh}}{\partial y} = 0,\label{eq:momVi}\\
&\frac{\partial w}{\partial t} + (U+u^\prime)\frac{\partial w}{\partial x} + (V+v^\prime)\frac{\partial w}{\partial y} +  w\frac{\partial w}{\partial z} + \frac{\partial p_{nh}}{\partial z} = 0,\label{eq:momWi}\\
&\frac{\partial \eta+\zeta^\prime}{\partial t} + \frac{\partial}{\partial x} \int\limits_{-d}^{\eta+\zeta^\prime} (U+u^\prime) \mathrm{d}z + \frac{\partial}{\partial y} \int\limits_{-d}^{\eta+\zeta^\prime} (V+v^\prime) \mathrm{d}z = 0.\label{eq:GCi}
\end{align}
\end{linenomath}

By taking the temporal average over the wave-motion scales and integrating the horizontal momentum equations over the vertical we obtain the following depth-averaged mean flow equations,
\begin{linenomath}
\begin{align}
&\frac{\partial U}{\partial x} + \frac{\partial V}{\partial y} = 0,\label{eq:LCm}\\
&\frac{\partial U}{\partial t} + U \frac{\partial U}{\partial x} + V \frac{\partial U}{\partial y}  + g \frac{\partial \eta}{\partial x} = - \int\limits_{-d}^{\eta} ( \overline{ u^\prime \frac{\partial u^\prime }{\partial x} } + \overline{ v^\prime \frac{\partial u^\prime }{\partial y} } ) \mathrm{d}z,\\
&\frac{\partial V}{\partial t} + U \frac{\partial V}{\partial x} + V \frac{\partial V}{\partial y}  + g \frac{\partial \eta}{\partial y} = -  \int\limits_{-d}^{\eta} ( \overline{ u^\prime \frac{\partial v^\prime }{\partial x} } + \overline{ v^\prime \frac{\partial v^\prime }{\partial y} } )  \mathrm{d}z,\\
&\frac{\partial\eta}{\partial t} + \frac{\partial \left(d+\eta\right) U }{\partial x} + \frac{\partial \left(d+\eta\right) V }{\partial y} = - \frac{\partial}{\partial x} \overline{ \int\limits_{-d}^{\eta+\zeta^\prime} u^\prime \mathrm{d}z }- \frac{\partial}{\partial y} \overline{ \int\limits_{-d}^{\eta+\zeta^\prime} v^\prime \mathrm{d}z }. \label{eq:GCm}
\end{align}
\end{linenomath}
In these equations, we can recognise the contribution to the radiation stress gradient from the orbital velocities (e.g., $\overline{ u^\prime \frac{\partial u^\prime }{\partial x} }$) and contributions in the global continuity equation that are related to stokes drift (i.e., the part of the integral above the wave trough in the right-hand-side of Eq. \ref{eq:GCm}). In the following we assume that waves do not influence the ambient currents, and neglect these contributions in the mean flow equations.

Subsequently, we derive a new set of wave equations by subtracting the mean equations \eqref{eq:LCm}-\eqref{eq:GCm} from the instantaneous equations \eqref{eq:LCi}-\eqref{eq:GCi},
\begin{linenomath}
\begin{align}
&\frac{\partial u^\prime}{\partial x} + \frac{\partial v^\prime}{\partial y}  + \frac{\partial w}{\partial z} = 0,\label{eq:LCf}\\
&\frac{\partial u^\prime}{\partial t}+u^\prime\frac{\partial  u^\prime}{\partial x} + v^\prime\frac{\partial u^\prime}{\partial y} + w \frac{\partial u^\prime}{\partial z} + g \frac{\partial \zeta^\prime}{\partial x} + \frac{\partial p_{nh}}{\partial x} = -( U \frac{\partial u^\prime}{\partial x} + u^\prime \frac{\partial U}{\partial x} + V \frac{\partial u^\prime}{\partial y} + v^\prime \frac{\partial U}{\partial y}),\\
&\frac{\partial v^\prime}{\partial t} + u^\prime\frac{\partial v^\prime}{\partial x} + v^\prime\frac{\partial  v^\prime}{\partial y} + w \frac{\partial v^\prime}{\partial z} + g \frac{\partial \zeta^\prime}{\partial y} + \frac{\partial p_{nh}}{\partial y} = -( U \frac{\partial v^\prime}{\partial x} + u^\prime \frac{\partial V}{\partial x} + V \frac{\partial v^\prime}{\partial y} + v^\prime \frac{\partial V}{\partial y}), \\
&\frac{\partial w}{\partial t} + u^\prime\frac{\partial w}{\partial x} + v^\prime\frac{\partial w}{\partial y} +  w\frac{\partial w}{\partial z} + \frac{\partial p_{nh}}{\partial z} = - ( U \frac{\partial w}{\partial x} + V \frac{\partial w}{\partial y} ) ,\\
&\frac{\partial \zeta^\prime}{\partial t} + \frac{\partial}{\partial x} \int\limits_{-d}^{\eta+\zeta^\prime} u^\prime \mathrm{d}z + \frac{\partial}{\partial y} \int\limits_{-d}^{\eta+\zeta^\prime} v^\prime \mathrm{d}z = - \frac{\partial \zeta^\prime U }{\partial x} - \frac{\partial \zeta^\prime V }{\partial y}. \label{eq:GCf}
\end{align}
\end{linenomath}
In the above set of equations, we can recognise the original set of equations (when dropping the prime superscripts) including several additional terms (on the right-hand-side) that account for the influence of a depth-uniform ambient current on the wave motions. We note that the influence of changes in the mean water level associated with the ambient current in the global continuity equation (i.e., the integral up to $\eta+\zeta^\prime$ in Eq. \eqref{eq:GCf}) can be straightforwardly incorporated by incorporating $\eta$ in the still water depth ($d=d+\eta$).

\subsection{Numerical implementation}
In the numerical implementation of the governing set of equations, the continuous description of time and horizontal dimensions are replaced by discrete approximations. In SWASH, the equations are discretised on regular or curvilinear grid for the horizontal dimensions and a terrain-following layering system for the vertical coordinate. A staggered grid arrangement is used to position the flow variables on the grid. Further details regarding the numerical implementation of the original set of equations can be found in several previous papers \citep[e.g.,][]{Stelling2003,Zijlema2005,Zijlema2011}, and will not be detailed here.

\begin{figure}[b!]
    \centering
    \includegraphics{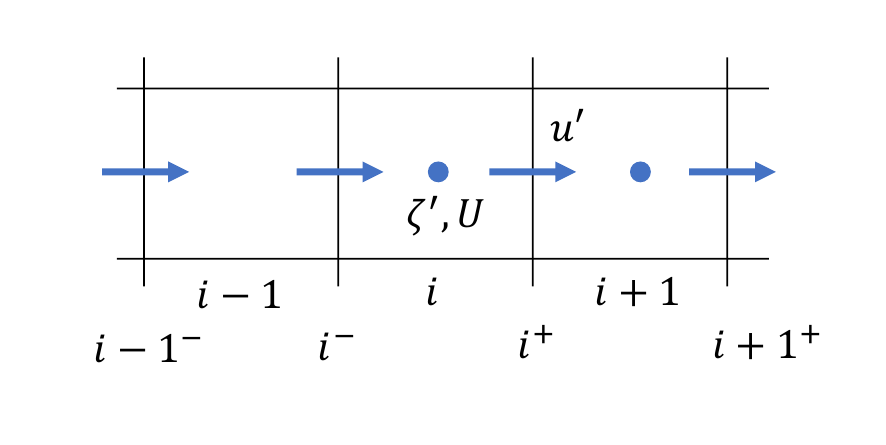}
    \caption{Illustration of the arrangement of the ambient velocity $U$ and wave-related variables $[\zeta,u]$ on the computational grid.}
    \label{fig:Grid}
\end{figure}

The flow velocities $[U,V]$ from the ambient current are positioned on the grid at the same location as the free-surface variable of the original set of equations $\zeta^\prime$  (i.e., at horizontal cell centres, see Fig. \ref{fig:Grid}). Linear interpolation is used to define the ambient current on the SWASH grid in the case that the ambient current is provided on a coarser grid. The numerical implementation of the additional terms is -- where possible -- based on the existing implementation of the advective terms. The terms in the horizontal momentum equations are discretised using the MacCormack predictor-corrector technique \citep{MacCormack1969} combined with flux limiters \cite[See][for more details]{Zijlema2011}. We use a flux limited first-order Euler scheme to discretise the terms in the vertical momentum equation. Finally, the terms in the global continuity equation are discretised using central differences and the Crank-Nicholson method.

\section{Linear properties of the model equations}\label{sec:LinProp}
We analysed the linear properties of the model equations by deriving the numerical linear dispersion relationship (see \ref{sec:A_LinAnalysis}) to verify that the model captures the effect of currents on waves. The numerical dispersion relationship derived from the extended model equations \eqref{eq:LCf}-\eqref{eq:GCf} provides a polynomial relationship $f_N$ between the absolute wave frequency $\omega$ (in the reference frame of a stationary observer) and the wavenumber $k$ for depth $d$ and current velocity $U$ depending on the number of layers $N$,
\begin{equation}
    \omega = f_N(k,d,U,N). \label{eq:numdisp}
\end{equation}

We compared linear wave properties based on this numerical dispersion relationship with the Doppler shifted dispersion relationship from linear theory \citep[e.g.,][]{Holthuijsen2007},
\begin{equation}
    \omega - k U = \sigma = \sqrt{ g k \tanh{k d} },\label{eq:dispDop}
\end{equation}
in which $\sigma$ is the intrinsic angular frequency (in the reference frame of an observer that is moving with the current). Based on this numerical and linear dispersion relationship, several wave properties can be derived. The relative group velocity (in a reference frame moving with the current) is given by $c_{g,r}=\frac{\partial\sigma}{\partial k}$, and the absolute group velocity (in the reference frame of a fixed observer) is $c_g = c_{g,r}+U$.

Furthermore, we also compared the numerical dispersion relationship of the extended model equations to the Doppler shifted numerical relationship of the original model equations,
\begin{equation}
    \omega - k U = \sigma = f_{N,U=0}(k,d,N),\label{eq:numdispDop}
\end{equation}
where $f_{N,U=0}$ is the numerical dispersion relationship in the absence of a current \citep{Smit2014}. This Doppler shifted numerical dispersion relationship provides the influence of a current on waves when the current is simulated as part of the model equations (e.g., by means of a pump system as described in \ref{sec:A_Pump}). Importantly, we found that all linear properties based on Eq. \eqref{eq:numdisp} (the numerical dispersion relationship of the extended model equations) and Eq. \eqref{eq:numdispDop} (the Doppler shifted numerical dispersion relationship) were identical. This confirms that the linear effect of current on waves can be captured by including additional terms in the model equations. In the remainder of this section, we therefore only compared linear wave properties based on the numerical dispersion relationship of the extended model equations \eqref{eq:numdisp} and the Doppler shifted dispersion relationship based on linear theory \eqref{eq:dispDop}.

\begin{figure}
    \centering
    \includegraphics[scale=0.9]{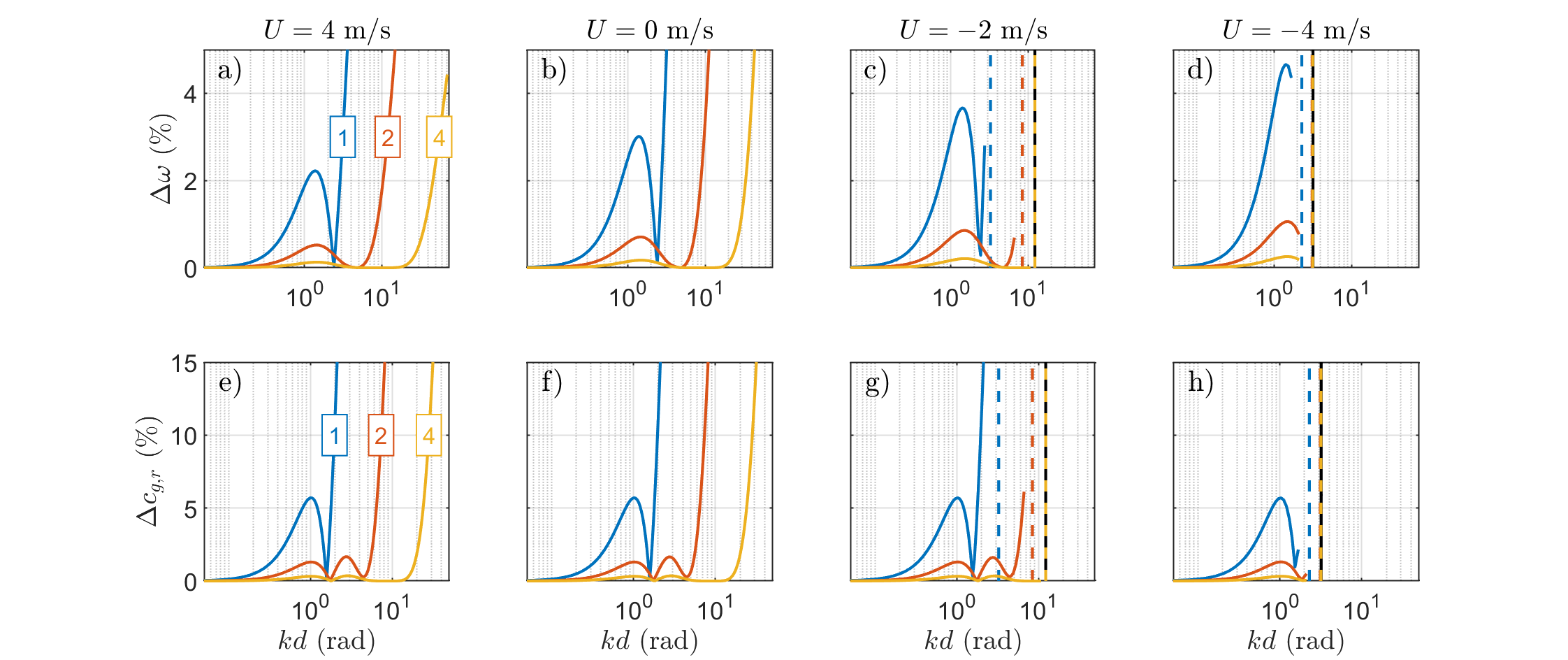}
    \caption{Absolute relative error in the absolute wave frequency $\omega$ (panel a-c) and relative group velocity $c_{g,r}=\frac{\partial\sigma}{\partial k}$ (panel d-f) as a function of the normalized water depth $kd$ for $U=[0,-2,-4]$ m/s (left to right panels, as indicated by the subplot titles) based on the numerical dispersion relationship of the $N$ layer system. Results are shown for $N=[1,2,4]$ layers. The vertical dashed lines indicate where blocking occurs, with the colors indicating the number of layers of the numerical dispersion relationship. The vertical black line indicates where blocking occurs according to the linear dispersion relionship.}
    \label{fig:LA_FD}
\end{figure}

\begin{figure}
    \centering
    \includegraphics{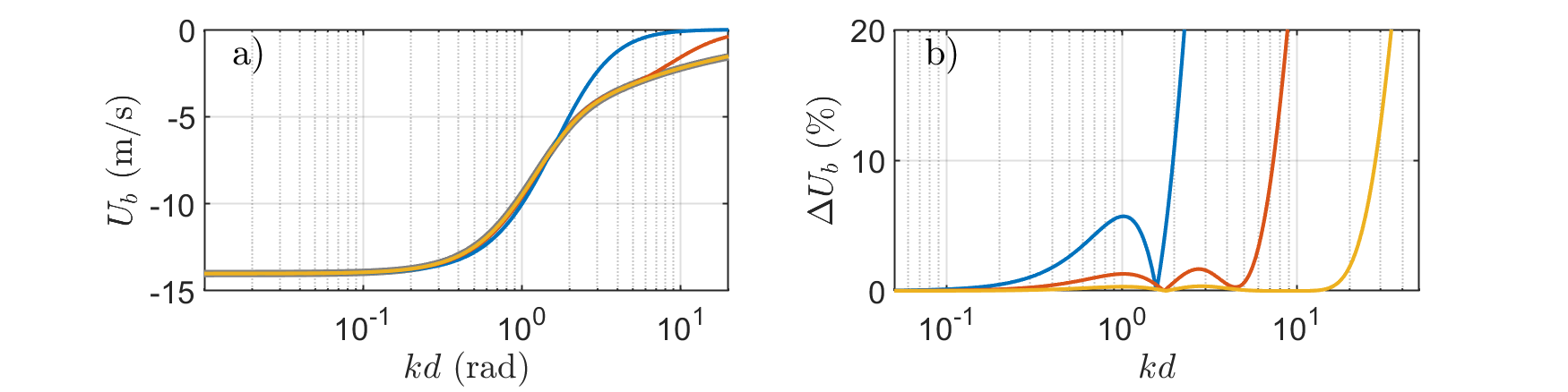}
    \caption{Panel a: Blocking current velocity $U_b$ (panel a) as a function of $kd$ based on the linear dispersion relationship (black line) and the numerical dispersion relationship for $N=[1,2,4]$ (blue, red, and yellow line, respectively). Panel b: Absolute relative error in $U_b$ from the numerical dispersion relationship for $N=[1,2,4]$ relative to the linear dispersion relationship as a function of $kd$.}
    \label{fig:LA_Ub}
\end{figure}

Assuming that the horizontal scales are sufficiently resolved, the dispersive property of the model depends on the number of vertical layers (Fig. \ref{fig:LA_FD}b). Introducing a current does not significantly affect the error in wave dispersion, as $\Delta\omega$ under currents is comparable to the case with $U=0$ m/s (compare Fig. \ref{fig:LA_FD}a,c,d with Fig. \ref{fig:LA_FD}b). Discrepancies in $c_{g,r}$ similarly depend on the number of layers and are not significantly affected by introducing a current (Fig. \ref{fig:LA_FD}e-f). When introducing an opposing current ($U<0$), no wave solution exists beyond a certain $kd$ as indicated by the vertical lines in Fig. \ref{fig:LA_FD}c-d and  \ref{fig:LA_FD}g-h. Here, waves are blocked as $c_g=0$. The $kd$ at which blocking occurs is sensitive to the number of layers, and is in better agreement with linear theory when a larger number of layers is used. This is further illustrated in Fig. \ref{fig:LA_Ub}, which shows the current velocity at which blocking occurs ($U_b$) as a function of $kd$ based on the linear and numerical dispersion relationship. With coarse vertical resolutions, waves are blocked on weaker opposing currents compared to linear theory. Increasing the number of vertical layers improves $U_b$, with errors in $U_b$ $<10\%$ for $kd<[2,7,30]$ in the case of $N=[1,2,4]$ layers, respectively. These findings show that the number of layers controls the accuracy with which the model recovers the linear wave properties in the presence of a current.

\section{Test Cases}\label{sec:TC}

\subsection{Linear waves on opposing and following currents}\label{sec:TC-LW}

To verify the numerical implementation of the additional terms in the governing equations, we compared model predictions of changes in the wavelength and wave amplitude due to a gradient in the current velocity to linear wave theory. As illustrated by the linear properties of the equations (Sec. \ref{sec:LinProp}), waves that travel over a current gradient experience a change in their kinematics. The wavelength decreases and the amplitude increases for waves on an opposing current and vice-versa on a following current. In this section, we verify if the developed model captures these changes to the wave field for linear waves that interact with opposing and following currents. We considered monochromatic waves with a height of $H=0.01$ m and wave periods $T=[5,10,15]$ s in water of constant depth $d=10$ m (corresponding to $kd=[1.7,0.7,0.4]$ in the absence of a current). A range of current velocities was simulated with $U$ ranging from -6 to 4 m/s with 0.25 m/s increments.

We compared the influence of the current on the wave height and the wavelength with linear wave theory. The change in wavelength and group velocity follows from the linear dispersion relationship \eqref{eq:dispDop}.
The change in wave height follows from the conservation of wave action,
\begin{equation}
    \frac{\partial}{\partial x} \frac{c_g E}{\sigma} = 0,
    \label{eq:action_balance}
\end{equation}
with the wave energy density $E$ of a monochromatic wave ($E=1/8 H^2$) and the absolute group velocity $c_g$ taken from linear theory (with $c_g=c_{g,r} + U$, and $c_{g,r}=\frac{\partial \sigma}{\partial k}$ obtained from the linear dispersion relationship).

\subsubsection{Model set-up}

To allow for the current effect on the waves to develop, the model setup included a transition region with a width of several wavelengths to gradually transition from no current to the respective maximum current velocity. The transition region had a width of $10 L_0$, and the region with maximum flow had a width of $10 L_0$ (with $L_0$ the wavelength in the absence of a current). These widths were found to be sufficient to allow for a gradual change in the wave dynamics, and provided a sufficiently large domain to determine the wave parameters in the presence of the current. Waves were generated at the left boundary with a wavemaker based on linear wave theory which was positioned $3 L_0$ away from the transition region. A sponge layer with a width of $5 L_0$ was positioned in front of the right boundary to absorb the waves and prevent any wave reflections. The sponge layer was positioned at a distance of $3 L_0$ from the transition region. The model was set-up with two layers in the vertical. The horizontal resolution and time-step were selected based on a sensitivity study (\ref{sec:A_Sens}): the horizontal grid resolution was set at $\Delta x=L_0/100$  and the time-step was set at $\Delta t=T/1000$  (with $T$ the incident wave period). The surface elevation $\zeta$ was outputted at all computational grid points for a duration of 5 wave periods after a spin-up time that ensured statistically stationary results inside the numerical domain.

We used zero-crossing analysis in the maximum current region to determine the wavelength in presence of a current. First, the surface elevation $\zeta$ was interpolated to a fine horizontal grid in the current region to allow for an accurate estimation of the wavelength independent of the grid resolution. The wavelength was subsequently computed from the zero-crossing analysis as the average wavelength over the current region and the output duration. We computed the wave height in the current region as $H =2 \sqrt{ 2 m_0 }$ (with the zeroth order moment $m_0$ computed as the standard deviation of the surface elevation $\zeta$). To gain insight in the spatial variation of $H$, we computed the mean, the maximum and minimum value of $H$ in the current region. Results were excluded when wave-blocking occurred in the model simulation. Wave blocking was recognised when the wave energy at the down-wave end of the domain (behind the current region) was $<1\%$ of the incident wave energy at the numerical wavemaker.

\subsubsection{Results}
\begin{figure}
    \centering
    \includegraphics{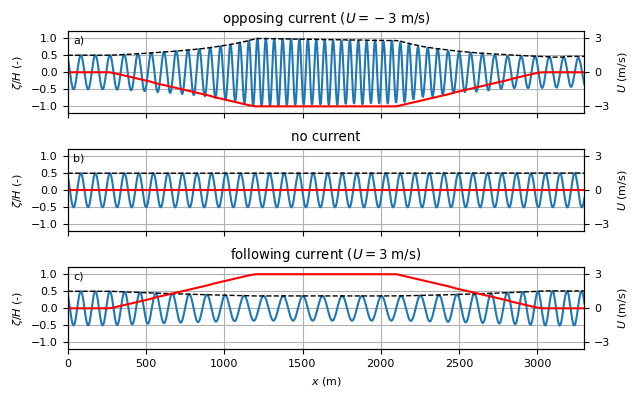}
    \caption{Snapshot of the modelled surface elevation (blue line, left axis) and ambient current velocity (red line, right axis) in the numerical domain for three different current velocities ($U=[-3,0,3]$ m/s) for a monochromatic wave with amplitude $a=0.01$ m and period $T=10$ s). The dashed black line indicates the envelope of the wave elevation, and the title of each panel indicates the respective current velocity.}
    \label{fig:LinearWave_IllustrativePlots}
\end{figure}

To illustrate the impact of the current on the wave field, Fig. \ref{fig:LinearWave_IllustrativePlots} shows an example of the surface elevation inside the model domain for three different current velocities. For these three cases, modelled changes to the surface elevation in opposing and following currents qualitatively agreed with the expected changes to the wave field. In an opposing current, the wavelength decreased and the wave height increased (Fig. \ref{fig:LinearWave_IllustrativePlots}a). In contrast, the wavelength increased and the wave height decreased for a following current (Fig. \ref{fig:LinearWave_IllustrativePlots}c). In all three illustrative cases, the wave signal at the downwave end of the flume ($x>3000$ m) was identical to the incident wave signal ($x=0$). This confirms that wave action is conserved in these simulations.

\begin{figure}
    \centering
    \includegraphics{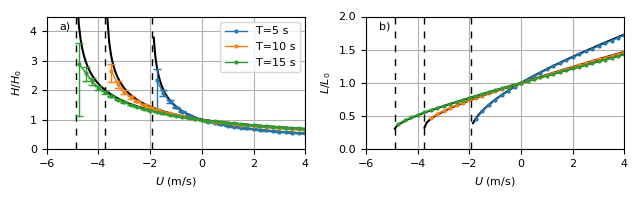}
    \caption{Normalized change to the wave height $H$ (panel a) and wavelength $L$ (panel b) as a function of the current velocity $U$ for small-amplitude monochromatic waves with $T=[5,10,15]$ s. The wave height and wavelength were normalized by the wave parameters in absence of a current (indicated by $[...]_0$). Converged model results of simulations (with $\Delta t=T/1000$ and $\Delta x=L/100$) are indicated by colored lines (see legend) and results from linear wave theory are indicated by the thick black line. In the left panel, the horizontal blue line with dotted markers indicates the average change to the simulated wave height $H$ in the current region and the vertical line with horizontal bars indicates the maximum and minimum simulated $H$ inside the current region. The dashed vertical black lines indicate the current velocity at which wave blocking occurs according to linear wave theory.}
    \label{fig:LinearWave_Comparison}
\end{figure}

To verify the model results quantitatively, we compared the change in the wave height and wavelength inside the current region with the results from linear wave theory (Fig. \ref{fig:LinearWave_Comparison}). For all three wave periods, linear wave theory predicted that the wave height and wavelength varied significantly for the considered range of current velocities (using Eq. \ref{eq:action_balance}). For opposing currents, the wave height $H$ increased and the wavelength $L$ decreased, and vice versa for following currents (as was visually observed in Fig. \ref{fig:LinearWave_IllustrativePlots}). Current induced changes to the wave field were larger for shorter wave periods, with wave blocking occurring for $T=[5,10,15]$ s at $U\approx[-1.92,-3.74,-4.87]$ m/s (indicated by the vertical black dashed line).

SWASH captured the changes to the wave height and wavelength for the range of ambient current velocities and the three wave periods (Fig. \ref{fig:LinearWave_Comparison}). This included the nonlinear dependence of $H$ and $L$ for $U<0$ m/s. Furthermore, the model captured blocking of waves for opposing currents that are stronger than the critical flow velocity of linear wave theory (indicated by the dashed black lines in Fig. \ref{fig:LinearWave_Comparison}). For all three wave periods, simulations with current velocities stronger than the theoretical blocking velocity showed a strong decay of the wave height down-wave of the blocking point (not shown). For simulations with $U$ close to but just weaker than the theoretical blocking velocity, dissipation of wave energy occurred in the model over the current region (visible as the difference between the vertical lines with horizontal bars in Fig. \ref{fig:LinearWave_Comparison}a, which indicates the maximum and minimum $H$ in the current region). In the absence of physical mechanisms for dissipation, this is likely related to numerical diffusion when the waves (with shorter lengths) propagate in the current region. For weaker $U$ this dissipation becomes smaller and the model results were in good agreement with linear theory. This numerical dissipation was found to be dependent on the horizontal grid resolution and time step, with improved agreement for strong $U$ for finer spatial and temporal resolutions (in accordance with the results in \ref{sec:A_Sens}). 

\subsection{Sheared current fields}\label{sec:TC-Akrish}
In coastal regions, spatially varying current fields exist (e.g., tidal currents) that can induce wave refraction and result in focal zones that give rise to wave interference patterns \citep[e.g.,][]{Yoon1989InteractionsWater,Akrish2020ModellingCurrents}. In this section, we verify the ability of the model to capture such wave patterns using two classical examples of wave-current interactions: the interactions of waves with a jet-like current and a vortex ring. Model results were compared with the spectral wave model SWAN \citep{Booij1999} extended with a quasi-coherent formulation that accounts for wave interference due to variable topography \citep{Smit2013a,Smit2015a} and currents \citep{Akrish2020ModellingCurrents}.

\subsubsection{Model set-up}
The model set-up was based on the work of \cite{Akrish2020ModellingCurrents}. The region of interest spanned a domain of $4\times4$ km. Two different simulations were considered, one with a jet-shaped and the other with a vortex-shaped current field, positioned along the central axis of the domain. The maximum velocities for the simulations were 0.38 m/s and 1.0 m/s, respectively \citep[refer to][for a mathemetical formulation of the current fields]{Akrish2020ModellingCurrents}. At the wavemaker positioned along the western boundary, a Gaussian shaped wave-spectrum in frequency and direction was forced with $H_s=1$ m, $T_p=20$ s and a standard deviation of 0.0015 Hz in frequency space and 1.78$^\circ$ in directional space. The waves had a mean direction of $\theta_0$=15$^\circ$ and $0^\circ$ (in Cartesian coordinates) for the jet and vortex current, respectively.

In the SWAN model, the physical domain was discretised with $\Delta x=\Delta y=50$ m. The spectral domain was discretised with 45 discrete frequencies that were logarithmically spaced between 0.005 and 0.085 Hz, and with a directional resolution of $2^\circ$ between -90 and 90$^\circ$. For the SWASH model, we extended the domain with a $500$ m wide sponge layer at the eastern side of the domain to prevent any wave reflections. The domain was discretised with a resolution of $\Delta x=2$ m and $\Delta y=$4 m (which resulted in $\approx$ 100 points per wavelength throughout the domain). The time step was set at $\Delta t=0.05$ s, equalling 300 points per wave period and resulting in $CFL\approx0.6$.

\begin{figure}[h!]
    \centering
    \includegraphics{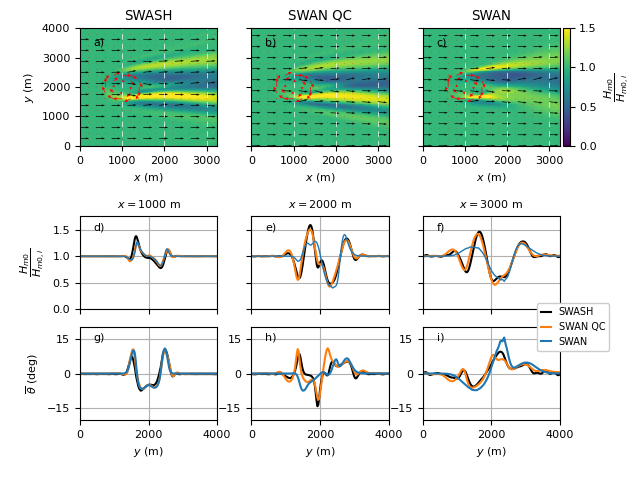}
    \caption{Changes to the significant wave height $H_{m0}$ and mean wave direction $\overline{\theta}$ due to a vortex ring current field. Panels a-c show a spatial overview of the significant wave height (colors) and mean wave direction (black arrows), with the red arrows indicating the ambient current field, for SWASH (panel a), SWAN including the Quasi-Coherent (QC) formulation (panel b) and default SWAN (panel c). Panels d-i show the wave height (d-f) and mean wave direction (g-i) along three alongshore transects predicted by SWASH (black lines), SWAN QC (orange lines) and default SWAN (blue lines).}
    \label{fig:Vortex}
\end{figure}

\begin{figure}[h!]
    \centering
    \includegraphics{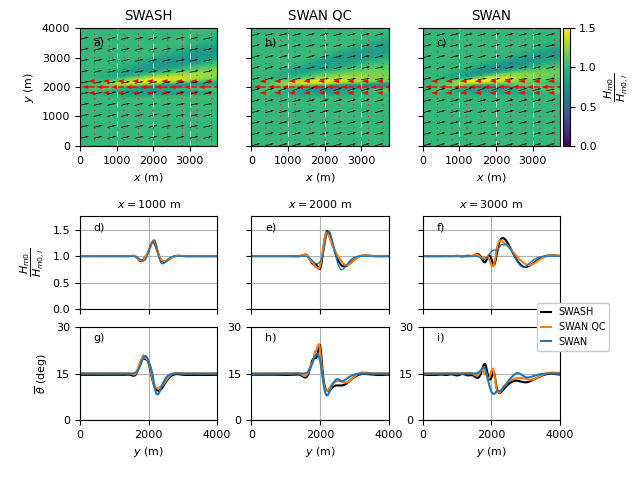}
    \caption{Changes to the significant wave height $H_{m0}$ and mean wave direction $\overline{\theta}$ due to a jet-like current field. Panels a-c show a spatial overview of the significant wave height (colors) and mean wave direction (black arrows), with the red arrows indicating the ambient current field, for SWASH (panel a), SWAN including the Quasi-Coherent (QC) formulation (panel b) and default SWAN (panel c). Panels d-i show the wave height (d-f) and mean wave direction (g-i) along three alongshore transects predicted by SWASH (black lines), SWAN QC (orange lines) and default SWAN (blue lines).}
    \label{fig:Jet}
\end{figure}

\subsubsection{Results}
Due to changes in wavelength induced by the current, waves were refracted by the vortex ring  (Fig. \ref{fig:Vortex}a-c and \ref{fig:Vortex}g-i). This current-induced refraction resulted in considerable variations in the significant wave height, with ridges of larger wave heights where waves focussed and depressions of lower wave heights where waves diverged (Fig. \ref{fig:Vortex}a-f). For this current field, quasi-coherent (QC) effects needed to be taken into account in SWAN to resolve the constructive and de-constructive wave interference that altered the wave field downstream of the vortex ring \citep[e.g.,][]{Akrish2020ModellingCurrents}. The bulk wave heights and mean wave directions predicted by the extended SWASH model were in satisfactory agreement with the results from the SWAN QC model throughout the domain.

Similarly, waves refract as they propagated into the jet-like current field, resulting in a change of mean wave direction (Fig. \ref{fig:Jet}g-i) and in regions with increased and decreased wave heights due to convergence/divergence of wave energy (Fig. \ref{fig:Jet}a-f). Similar to the vortex ring, quasi-coherent effects need to be incorporated in SWAN to account for the constructive and de-constructive wave interference that altered the wave field, although this effect was smaller compared to the vortex ring. In general, the SWASH predictions were in good agreement with SWAN QC. The results of this test case, and the vortex ring, illustrate that SWASH including the additional terms in the model equations is able to capture the effect of current-induced refraction on the wave propagation and the resulting spatial variability in the wave field.

\subsection{Wave blocking, reflections and breaking on opposing currents}\label{sec:TC-C99}
\begin{figure}
    \centering
    \includegraphics{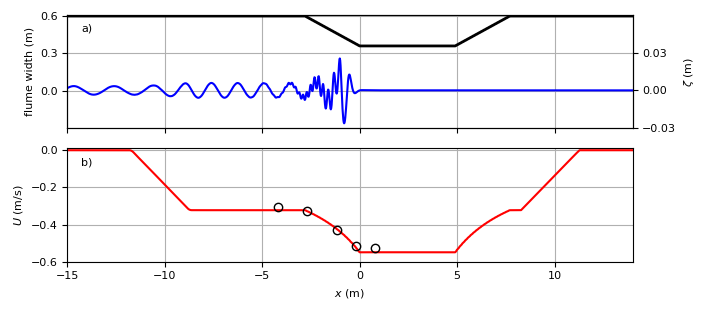}
    \caption{Overview of the numerical setup of the \citet{Chawla1999ExperimentalCurrents} flume experiment. The top panel (a) shows the flume width (black line, left axis) and a snapshot of the modelled free-surface elevation for test case 1 (blue line, right axis). The bottom panel (b) shows the modelled (red line) and measured (black markers) current velocity (in the absence of waves).}
    \label{fig:C99_U}
\end{figure}

As a final test case, we compare model predictions with the laboratory experiment of \cite{Chawla1999ExperimentalCurrents,Chawla2002MonochromaticPoints} that considered wave blocking on opposing currents. The flume had a length of 30 m, a width of 0.6 m and still water depth of 0.5 m, with a pump system to generate a recirculating current (with a discharge of 0.095 m$^3$/s) and a perforated wavemaker to generate waves on the current. A spatially varying current was generated by means of a false wall constricting the width of the flume, with a minimal width of 0.36 m (see black line in Fig. \ref{fig:C99_U}a). Blocking of waves occurred close to the start of this narrow part of the flume.

The experiments with monochromatic waves considered a total of 23 test conditions that included 3 different incident wave periods ($T=[1.2,1.3,1.4]$ s) for a range of wave heights ($H=0.012-0.14$ m). For the low amplitude and low period waves, waves reflected with negligible transmission of wave energy beyond the blocking point. For increasing wave heights, waves started breaking at the blocking point of linear theory combined with increased transmission of wave energy beyond this theoretical blocking point. In this paper, we considered 4 out of the 23 test cases: the largest and smallest wave height of both the smallest and largest wave period (see Table \ref{tab:Chawla}). For case R1 and R11 waves reflected at the blocking point, whereas waves were breaking and wave energy was transmitted beyond the theoretical blocking point for case B6 and B18.

We compared model predictions with these laboratory observations for these 4 test cases. Furthermore, we also computed the wave height transformation based on conservation of wave action.
\begin{equation}
    \frac{\partial}{\partial x}\frac{c_g E}{\sigma} = 0,
\end{equation}
Conservation of wave action is computed based on the linear dispersion relationship (similar to Sec. \ref{sec:TC-LW}) and also based on the nonlinear dispersion relationship from 2nd order Stokes theory \citep[e.g.,][]{Dean1991WaterScientists}. This nonlinear dispersion relationship accounts for the effect of amplitude dispersion, which was found to be important for these laboratory experiments \citep{Chawla2002MonochromaticPoints}.

\begin{table}[]
    \centering
    \begin{tabular}{c|c|c|c|c|c|c}
            & $H$ (m) & $T$ (s) & $U_b$ (m/s) & $kd$ ($U=0$ m/s) & $kd$ ($U=-0.32$ m/s) & $kd$ ($U=U_b$ m/s) \\ \hline
    R1 & 0.012 & 1.2 & -0.47 & 1.53 & 2.36 & 5.59\\
    B6 & 0.126 & 1.2 & -0.47 &  1.53 & 2.36 & 5.59\\
    R11  & 0.015 & 1.4 & -0.55 &  1.22 & 1.69 & 4.14 \\
    B18 & 0.141 & 1.4 & -0.55 &  1.22 & 1.69 & 4.14 \\
    \end{tabular}
    \caption{Experimental conditions (wave height $H$, wave period $T$, theoretical blocking velocity $U_b$, and normalized water depth $kd$ for three current velocities) of the four test cases of the \cite{Chawla1999ExperimentalCurrents} flume experiment that were considered in this paper.  The wave height and wave period were measured at the first wave gauge inside the flume, at a distance of 4.2 m (case 1 and 3) and 5.2 m downstream (case 2 and 4) of the start of the narrow flume section. The blocking velocity was computed based on the linear dispersion relationship. The normalized water depth based on linear wave theory is provided in the absence of the current, for $U=-0.32$ m/s, and at the theoretical blocking velocity $U_b$.}
    \label{tab:Chawla}
\end{table}

\subsubsection{Model setup}
We used a curvilinear grid with a constant streamwise resolution but varying alongshore width and resolution to replicate the flume in the numerical model. Based on the sensitivity study for linear waves (\ref{sec:A_Sens}), the horizontal grid resolution in streamwise direction was set to ensure at least 100 points per wavelength (in the absence of a current). This resulted in a total of 1500 cells in the streamwise direction. We used 3 cells in the spanwise direction to reduce computational overhead. This implies that spanwise effects were not included in the modelling, such as the sidewall boundary layers that were observed in the flume \citep{Chawla2002MonochromaticPoints}. To investigate the influence of the vertical resolution, simulations were run with $[2,4,20]$ layers.
The model time step was set at a value that corresponds to $CFL\approx 0.4$ resulting in about 250-500 points per wave period, which was found to be sufficiently fine for these test conditions.
Waves were generated based on linear wave theory at $x=-15$ m (in the absence of a current), with the incident wave height calculated from conservation of wave action (Eq. (\ref{eq:action_balance})) based on the measured wave height at the first wave gauge (located at $x\approx-5$ m). A sponge layer with a width of at least three wavelengths was positioned at the end of the flume to prevent wave reflections.

We conducted two sets of simulations to replicate the four test cases. In the first set, which serves as a benchmark for the proposed model extension, the waves and current were modelled simultaneously through the original set of model equations. A re-circulating current was generated through modifying the kinematic boundary condition at the bottom (see \ref{sec:A_Pump}). The resulting discharge that is imposed at the bottom replicates a pump system through which a volume of water is pumped into the domain at one end of the flume and is taken out at the other end of the flume. In this manner, a current was generated inside the numerical flume. In all simulations with this pump system, the discharge was set at $Q=0.095$ m$^{3}$/s based on \cite{Chawla2002MonochromaticPoints}. With this model set-up, the modelled depth-averaged current field was in good agreement with observations taken in the flume for a reference case excluding waves (Fig. \ref{fig:C99_U}b). In the second set of simulations, we account for the current through the additional terms in the equations that were derived in Section \ref{sec:NumMethod}. The ambient current velocities were obtained from the simulation with the pump system without waves (Fig. \ref{fig:C99_U}b). In the following, we refer to the simulations with the additional terms to model the influence of the current on waves as an Ambient Current (AC) simulation, and we refer to the benchmark simulations as a Pump simulation.

Non-hydrostatic models like SWASH inherently account for the dissipation by breaking waves but require high vertical resolutions to capture the onset of wave breaking correctly \citep[e.g,.][]{Smit2013}. To capture the onset of breaking with coarse resolutions, \citet{Smit2013} introduced the Hydrostatic Front Approximation (HFA) that neglects the non-hydrostatic pressure locally to trigger wave breaking (i.e., switching to the Non-Linear Shallow Water Equations, NSLWE). However, numerical instabilities developed in all 2-layer simulations with HFA. We believe this is related to the normalized water depth of the waves at breaking. For depth-induced wave breaking in the absence of currents (for which the HFA is normally applied), the normalized water depth is relatively low at breaking ($kd<1$), resulting in a relatively small non-hydrostatic pressure contribution. In the wave-current simulations of this test case, the normalized water depth is relatively large ($kd>4$ near wave blocking, see Table \ref{tab:Chawla}). As a results, the contribution from the non-hydrostatic pressure is relatively large at the location of incipient wave breaking. Excluding a relatively large contribution from the non-hydrostatic pressure likely resulted in numerical instabilities and caused the model to crash. As a result, the HFA approach cannot be used to improve the model predictions of the 2-layer model in the case of breaking waves on a strong opposing current. In the following, we therefore only show results for 2-layer simulations excluding HFA.


\subsubsection{Results - wave reflections}

\begin{figure}[t!]
    \centering
    \includegraphics[scale=0.8]{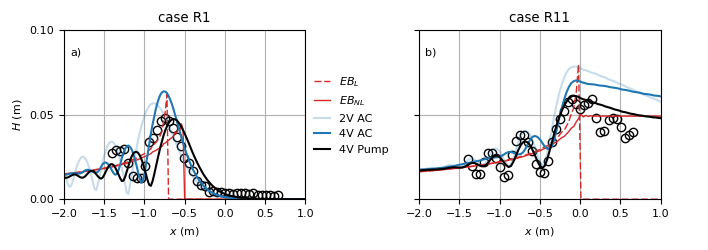}
    \caption{Comparisons between the measured and modelled wave height for test cases R1 and R11 of the wave-current flume experiment of \cite{Chawla1999ExperimentalCurrents}. The black circle markers indicate the experimental observations, and the coloured lines indicate the model predictions (light and dark blue, 2 and 4-layer simulations with AC (Ambient Current), respectively; black, benchmark 4-layer simulation with pump system). The thin red lines show the results from an energy balance based on conservation of wave action using the linear dispersion relationship (dashed red line) and the 2nd order Stokes dispersion relationship (full red line).}
    \label{fig:C99-R}
\end{figure}

For the wave condition with the smallest wave height and wave period (case R1, Table \ref{tab:Chawla}), waves reflected at the blocking point, resulting in a nodal pattern in the wave height $H$ and negligible transmission of wave energy for $x>0$ m (Fig. \ref{fig:C99-R}a). An energy balance based on conservation of wave action (eq. \eqref{eq:action_balance}) provided a reasonable good description of the location of wave blocking. Differences between the energy balance with the linear dispersion relationship and 2nd order Stokes dispersion relationship were generally small except near the blocking location, where the blocking location is spatially shifted by approximately 0.25 m when accounting for amplitude dispersion.

In Fig. \ref{fig:C99-R}a, we compare both results of the simulations with an Ambient Current (AC) and of a benchmark simulation in which the current is included through a re-circulating pump. Both model setups (AC and Pump) reproduced this blocking and reflection of waves as the simulations captured the nodal pattern in the wave height for $x<0$ m and wave energy was not transmitted beyond $x=0$ m (Fig. \ref{fig:C99-R}a). Model simulations were found to be sensitive to the number of layers, and were approximately converged for 4 layers (as illustrated by the results of the AC simulations). The nodal structure in $H$ was stronger and spatially shifted towards the wavemaker using two vertical layers with both the AC (Fig. \ref{fig:C99-R}a) and Pump setup (not shown). This indicates that reflections were stronger and blocking occurred at a weaker opposing current velocity when using this coarsest vertical resolution. Increasing the vertical resolution improved the results of the AC simulation, although $H$ was over predicted at blocking compared to the measurements and the benchmark simulation. Results of the 4-layer benchmark Pump simulation were in good agreement with the measurements, apart from a slight spatial shift (approx. $0.15$m) of the blocking location and nodal pattern.

For test case R11, blocking was expected at $x\approx 0$ m based on the energy balance with linear dispersion (Fig. \ref{fig:C99-R}b). In contrast, the energy balance with 2nd order dispersion predicted no blocking but transmission of energy for $x>0$m. In the laboratory, partial reflections occurred at the blocking point with partial transmission of energy for $x>0$. Both the AC and Pump simulations captured these patterns. Similar to case R1, simulations approximately converged when 4 layers were used. The 4-layer benchmark simulation was in best agreement with the observations, and captured both the spatial variability and magnitude of $H$. The 4-layer AC simulations overpredicted $H$ near the linear blocking point for $x>0$ m (similar to test case R1) and predicted weaker reflections resulting in a less pronounced nodal pattern for $x<0$ m.

These results show that the proposed extension of the model equations captured the overall patterns in the wave height that was observed in the laboratory and simulated by the benchmark model. Model results of both the AC and Pump simulations were found to be sensitive to the number of layers, indicating that the dispersive properties of the model affected the location of blocking and controlled the magnitude of wave reflections. Discrepancies in the blocking location at coarse vertical resolutions were larger for case R1 (with a smaller wave period and thus larger $kd$ compared to R11). This is consistent with the expected response based on the numerical dispersion relationship (Fig. \ref{fig:LA_Ub}): the relative absolute error in $U_b$ compared to linear theory was 1.49$\%$ and 0.42$\%$ for R1 and R11 when using 2 layers, respectively, and $<0.25\%$ when using 4 layers.


\subsubsection{Results - wave breaking}

\begin{figure}[b!]
    \centering
    \includegraphics[scale=0.8]{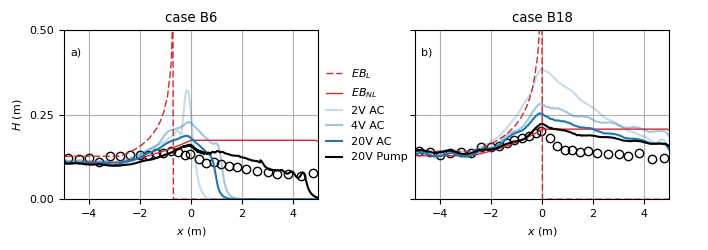}
    \caption{Comparisons between the measured and modelled (20-layer simulations) wave height for test cases B6 and B18 of the wave-current flume experiment of \cite{Chawla1999ExperimentalCurrents}. The black circle markers indicate the experimental observations, and the coloured lines indicate the model predictions (light to dark blue, 2, 4 and 20-layer simulations with AC (Ambient Current); black, benchmark 20-layer simulations with pump system).
    The thin red lines show the results from an energy balance based on conservation of wave action using the linear dispersion relationship (dashed red line) and the 2nd order Stokes dispersion relationship (full red line).}
    \label{fig:C99-B}
\end{figure}

For larger incident wave heights (case B6 and B18), wave breaking on the opposing current was observed during the experiment in the narrow region of the flume ($x=0-5$ m) and wave energy was transmitted beyond the blocking point from linear theory (Fig. \ref{fig:C99-B}). For case B6, the 2-layer AC simulation did not capture the transmission of wave energy beyond the blocking point, but showed signs of wave reflections near $x=0$ m, resulting in an over prediction of the wave height for $-2<x<0$ m (Fig. \ref{fig:C99-B}a). Similar results were observed for the simulation with a Pump system (not shown). These results indicate that the 2-layer simulations failed to capture the breaking of waves and transmission of energy beyond the linear blocking point for this particular test case. Increasing the number of vertical layers significantly improved the model results (as indicated by the 4 and 20-layer AC simulations, Fig. \ref{fig:C99-B}a). In particular, the 20V Pump benchmark simulation captured $H$ throughout most of the domain, including the transmission of energy beyond the linear blocking point and the gradual decay of $H$ for $x>0$ m. The 20V AC simulation also captured part of this wave transmission but only up to $x\approx 1$ m, and over predicted $H$ near $x=0$ m.

For case B18, with a larger incident wave height and period, wave energy was transmitted beyond the linear blocking point for the 2-layer simulation with no sign of wave reflections (Fig. \ref{fig:C99-B}b). However, $H$ was over predicted for $x>-2$m. Simulations with the Pump provided similar results (not shown). Increasing the number of vertical layers significantly improved the model results, and 20V AC simulations agreed well with 20V Pump simulations apart from a slight over prediction for $x>-1$ m. Both 20V models were also in satisfactory agreement with the observations, apart from an over prediction of $H$ for $x>0$ m.

For the test cases with breaking waves, the model predictions were found to be sensitive to the vertical resolution. A relatively fine vertical resolution was found to be required to capture changes in the wave height. For case B6, a fine vertical resolution was required to prevent wave-reflections at the blocking point and to capture (part of) the transmission of energy for $x>0$. In contrast, wave energy was transmitted beyond the linear blocking point at coarse resolutions for case B18. For this test case, higher vertical resolutions were required to better capture the shoaling of waves on the opposing current. The shoaling in 2-layer simulations was similar to the linear energy balance, whereas shoaling in the case of more vertical layers was comparable to the nonlinear energy balance and the measurements. This suggests that for B18 a higher vertical resolution is required to capture the effect of (nonlinear) amplitude dispersion.

\section{Discussion}\label{sec:Discussion}
The results of this work demonstrated that the extended SWASH model was able to capture the dominant effects of currents on waves. Comparisons with linear theory and a spectral wave model showed that the model captured current-induced changes to the wave amplitude and length, and current-induced refraction. Comparisons with the laboratory experiment of \citet{Chawla1999ExperimentalCurrents,Chawla2002MonochromaticPoints} showed that the model reproduced the (partial) reflection of monochromatic waves on an opposing current near the blocking point in the case of small amplitude waves, and (partial) transmission and wave breaking in the case of larger amplitude waves. For these challenging test cases, model results were found to be sensitive to the number of vertical layers. In particular, a fine vertical resolution was required to capture the nonlinear shoaling, breaking and (partial) transmission of the large amplitude waves on the opposing current. Importantly, the results from the extended SWASH model were generally in good agreement with fully resolved benchmark simulations that intrinsically accounted for the wave-current interactions. This indicates that additional physics in the fully resolved SWASH model (e.g., vertical variations in the ambient flow, and the influence of waves on the ambient currents) did not significantly affect the wave dynamics in these test cases.

Instead, this indicates that model-data discrepancies were largely inherited from the fully resolved model. For example, these could be related to the exclusion of span-wise flow effects, and shortcomings in the turbulence modelling (e.g., no wave breaking induced turbulence at the free-surface, incomplete description of turbulent boundary layers). To our knowledge, current state-of-the-art CFD models such as RANS and SPH-type models have not been widely used to simulate these nor similar laboratory experiments that consider such complex wave-current interactions. Only a few authors have used CFD for selected cases of laboratory experiments \citep[e.g.,][]{Olabarrieta2010,Teles2013NumericalScale,Chen2018CharacteristicsCurrents,Yao2023ACurrent} but not for a wide variety of conditions such as the reflective and breaking cases that were considered in this work. As such, we currently lack a clear benchmark that indicates how accurate fully resolved 3D models including more sophisticated turbulence models can capture wave-current interactions.

\section{Conclusions}\label{sec:Conclusions}
This study has demonstrated that the non-hydrostatic modelling approach can be extended to account for the effect of depth-uniform currents on the wave dynamics. By introducing a separation of scales and assuming vertically uniform mean currents, additional terms were derived that account for changes in the wave properties in the presence of spatially varying currents. These additional terms were included in the open-source SWASH model.

A linear analysis of the model equations confirmed that the proposed model extension resolves the effect of currents on the linear wave properties (e.g., change in wavelength and group velocity). Comparisons of model predictions with linear wave theory further verified the numerical implementation. The extended SWASH model captured changes in the wavelength and amplitude in the presence of opposing and following currents for small amplitude waves. As a next step, we validated the model for more complex spatially varying flow fields: a vortex ring and a jet-like current. SWASH predictions were compared with the spectral wave model SWAN, including the Quasi-Coherent formulation to account for constructive and de-constructive wave interference effects. Comparisons of bulk wave parameters (significant wave height and mean wave direction) showed that the extended SWASH model was able to account for the current-induced refraction of both flow fields, and the resulting spatial variability in the wave height.

Finally, we compared model predictions with a flume experiment that considered blocking and breaking of monochromatic waves on a strong opposing current. Although the model tended to overpredict the wave height, it was able to reproduce reflections of small amplitude waves, and breaking of larger amplitude waves. For breaking waves, model results were improved by increasing the vertical resolution (from 2 to 20 layers). Results of the newly derived model were generally consistent with fully resolved SWASH simulations (in which a recirculating current was included through an inflow and outflow boundary at the bottom). This indicates that model-data discrepancies were largely inherited from the fully-resolved model and not introduced by missing physics in the extended model (e.g., no vertical variation of the ambient current, and no effect of waves on the ambient current).

The findings of this work thereby demonstrated that phase-resolving models can be extended with additional terms to account for the major effect of ambient depth-uniform currents on the wave dynamics. This will allow models like SWASH to more accurately and efficiently simulate the wave dynamics in coastal environments where tidal and/or wind-driven currents are present.


\appendix
\setcounter{figure}{0}

\section{Sensitivity study}\label{sec:A_Sens}
The behaviour of the SWASH model was found to be sensitive to the horizontal grid resolution $\Delta x$ and the time-step $\Delta t$. To illustrate the sensitivity to the grid resolution, we consider a set of simulations of a $T=10$ s monochromatic wave on a $U=[-3,-1]$ m/s current for a range of horizontal grid and temporal resolutions. To study the influence of $\Delta t$ and $\Delta x$ separately, the first set considers simulations with fixed $\Delta x=L_0/60$ for a range of $\Delta t$, and the second set corresponds to several simulations with fixed $\Delta t=T/1000$ but for a range of $\Delta x$.

\begin{figure}[!t]
    \centering
    \includegraphics{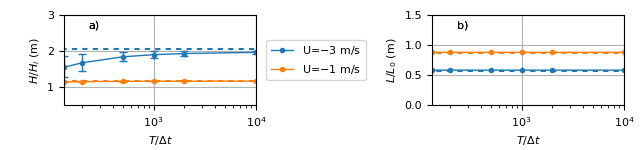}
    \newline
    \includegraphics{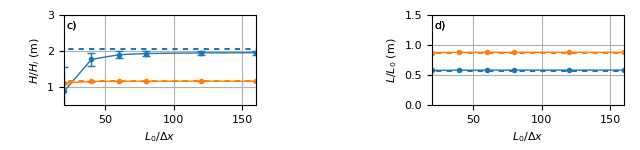}
    \caption{Changes to the height (panel a and c) and length (panel b and d) of a monochromatic wave ($T=10$ s) on an opposing current ($U=[-3,-1]$ m/s) as a function of the temporal resolution with a fixed grid resolution $\Delta x/L_0=60$ (panel a-b) and as a function of the horizontal grid resolution with a fixed temporal resolution $\Delta t=T/1000$ (panel c-d). The full lines indicate the SWASH results and the dashed lines indicate the results according to linear wave theory. Results for $U=-3$ m/s are printed in blue and results for $U=-1$ m/s in orange. For SWASH, the horizontal line with marker indicates the average change to the simulated wave height $H$ in the current region, and the vertical lines with horizontal endings indicate the maximum and minimum $H$ in the current region. The wave height and length are normalized by the incident wave height and length, respectively.}
    \label{fig:LinearWave_Sensitivity}
\end{figure}

Changes to the wavelength $L$ were not sensitive to either $\Delta x$ and $\Delta t$. On the other hand, changes to the wave height $H$ were sensitive to the model settings. The sensitivity was larger for the stronger current velocity. Modelled changes to $H$ were less sensitive to the horizontal grid resolution, except for coarse resolutions ($\Delta x/L_0<40$), with relatively weak improvement for $\Delta x/L_0\leq40$ (Fig. \ref{fig:LinearWave_Sensitivity}). Modelled changes to $H$ were sensitive to the time-step, especially for $U=-3$ m/s. For this current velocity at larger time-steps, significant dissipation of wave energy occurred in the current region (as illustrated by the vertical lines in Fig. \ref{fig:LinearWave_Sensitivity} at smaller $\Delta t / T$). For finer temporal resolutions, this non-physical dissipation reduced and model results approximately converged to the solution of linear wave theory. This sensitivity to the horizontal grid and temporal resolution was primarily significant for strong opposing currents relative to the wave group velocity. For following currents and weak opposing currents the model results were not sensitive to $\Delta x$ and $\Delta t$ (as illustrated by the results for $U=-1$ m/s). Based on this sensitivity study, the optimal horizontal grid and temporal resolution for which model predictions were sufficiently converged was concluded to be $\Delta x=L_0/100$ and $\Delta t=T/1000$.

\section{Re-circulating current}\label{sec:A_Pump}
To generate a re-circulating current in the model, we impose an inward and outward flux at the bottom at either side of the model domain. For this purpose, we have adopted the kinematic boundary condition as follows,
\begin{linenomath}
\begin{equation}
    w_{z=-d}= - u \frac{\partial d}{\partial x} - v \frac{\partial d}{\partial y} \pm f_s \frac{P}{W},
\end{equation}
\end{linenomath}
where $P$ is a discharge and $W$ is the width of the region where the discharge is specified. By introducing an equal discharge of opposing sign in a region at either side of the numerical domain, a recirculating current is generated inside the domain. To reduce the spin-up time, we use a smoothing function $f_s$ to gradually ramp up the discharge from 0 to $P$. The smoothing function is defined as,
\begin{linenomath}
\begin{equation}
    f_s = 0.5 \; ( 1 + \tanh( \frac{t}{T_S} - 3 ) ),
\end{equation}
\end{linenomath}
where $T_S$ is the smoothing period of the pump (taken as $T_S=15$ s in the simulations of this work).

\section{Linear semi-discrete analysis of the model equations}\label{sec:A_LinAnalysis}

The numerical dispersion relationship can be derived from the linearized and semi-discretized set of model equations \citep[e.g.,][]{Cui2014,Bai2013,Smit2014}. Based on \cite{Smit2014}, the linearized and semi-discretized SWASH equations extended with the additional terms for the wave-current interactions (on the right hand side) for $N$ vertical layers reads,
\begin{align}
    &\frac{\partial u^\prime_{n-\frac{1}{2}} }{\partial t} + g \frac{ \partial \zeta^\prime }{\partial x} + \frac{1}{2} \frac{\partial p_{nh,n} }{\partial x} + \frac{1}{2} \frac{\partial p_{nh,n-1} }{\partial x} = - U \frac{\partial u^\prime_{n-\frac{1}{2}}}{\partial x}, \;\;\;\; \rm{for} \; n= 1...N,\\
    &\frac{ \partial w_n + w_{n-1} }{\partial t} + 2 \frac{p_{nh,n}-p_{nh,n-1}}{\Delta z} = - U \frac{ \partial w_n + w_{n-1}}{\partial x}, \;\;\;\; \rm{for} \; n= 1...N,\\
    &\frac{ \partial u^\prime_{n-\frac{1}{2}} }{\partial x} + \frac{ w_{n} - w_{n-1} }{\Delta z} = 0, \;\;\;\; \rm{for} \; n= 1...N,\\
    &\frac{ \partial \zeta^\prime }{\partial t} + \Delta z \sum\limits_{n=1}^{N} \frac{ \partial u^\prime_{n-\frac{1}{2}} }{\partial x} = -U \frac{\zeta^\prime}{\partial x}.
\end{align}
The flow variables in the above set of equations are located on a staggered grid, with $u^\prime$ located in a cell center ($n-\frac{1}{2}$) and $w$ and $p_{nh}$ at a vertical cell face ($n$).
Assuming a horizontal bottom ($w_0$=0) and considering the initial value problem in an infinite domain (with $\Delta z=d/N$), we assume that the flow variables have a solution of the form $y=\hat{y} \rm{exp}(ikx-i\omega t)$ (where $\hat{y}$ is the complex amplitude of a flow variable, $k$ is the wavenumber and $\omega$ the absolute wave frequency). Substituting this into the above set of equations for each variable results in a system of equations of the form $A\hat{y}=0$. The numerical dispersion relationship can subsequently found from $\rm{Det}(A)=0$ using symbolic algebra software. With the addition of an ambient current $U$, the numerical dispersion relationship provides a relationship between $\omega$ and $k$ in the presence of a current with velocity $U$ for $N$ vertical layers. The relative group velocity can be found from the numerical dispersion relationship as $c_{g,r}=\frac{\partial \sigma}{\partial k}$ for an arbitrary current velocity $U$ (with $\omega=\sigma+kU$).
\clearpage
\bibliographystyle{elsarticle-harv}


\end{document}